\documentclass[twocolumn,aps]{revtex4}
\usepackage{graphicx}
\usepackage{helvet}

\def\eq#1{Eq.~(\ref{#1})}
\def\fig#1{Fig.~\ref{#1}}

\begin{document}

\title{Cell organization in soft media due to active mechanosensing}

\author{I. B. Bischofs}
\author{U. S. Schwarz}
\email[To whom correspondence should be addressed. Email: ]{Ulrich.Schwarz@mpikg-golm.mpg.de}
\affiliation{Max Planck Institute of Colloids and Interfaces, 14424 Potsdam, Germany}

\begin{abstract}
Adhering cells actively probe the mechanical properties of their
environment and use the resulting information to position and orient
themselves. We show that a large body of experimental observations can
be consistently explained from one unifying principle, namely that
cells strengthen contacts and cytoskeleton in the direction of large
effective stiffness. Using linear elasticity theory to model the
extracellular environment, we calculate optimal cell organization for
several situations of interest and find excellent agreement with
experiments for fibroblasts, both on elastic substrates and in
collagen gels: cells orient in the direction of external tensile
strain, they orient parallel and normal to free and clamped surfaces,
respectively, and they interact elastically to form strings. Our
method can be applied for rational design of tissue
equivalents. Moreover our results indicate that the concept of contact
guidance has to be reevaluated. We also suggest that cell-matrix
contacts are upregulated by large effective stiffness in the
environment because in this way, build-up of force is more efficient.
\end{abstract}

\maketitle

The mechanical activity of adherent cells usually is attributed to
their physiological function. For example, fibroblasts are believed to
maintain the structural integrity of connective tissue and to
participate in wound healing by actively pulling on their environment.
During recent years, it has become clear that there is another
important role for mechanical activity of adherent cells: by pulling
on their environment, cells can actively sense its mechanical
properties and react to it in a specific way
\cite{c:galb98,c:huan99,c:geig02}. Harris and coworkers observed
surprisingly large tension fields for fibroblasts on elastic
substrates, which induce mechanical activity of other cells, even when
located at considerable distance \cite{c:harr80}.  When plated on
elastic substrates of increased rigidity, many cell types show
increased spreading and better developed stress fibers and focal
adhesions \cite{c:pelh97}.  Fibroblasts on elastic substrates orient
in the direction of tensile strain \cite{c:hast83} and locomote in
favor of regions of larger rigidity or tensile strain
\cite{c:lo00}.  The same response has been reported
for vascular smooth muscle cells on rigidity gradients
\cite{c:wong03}. Similar observations have been reported numerous
times also for tissue cells in hydrogels. For fibroblasts in collagen
gels, Bell and coworkers not only found that traction considerably
contracts the gel, but also reported orientational effects: cells
align along the direction of pull between fixed points and parallel to
free surfaces \cite{c:bell79}. When a collagen gel is stretched
uniaxially, cells orient in the direction of principal strain
\cite{c:east98}. Moreover, cells align in a nose-to-tail
configuration, thus forming strings running in parallel to the
direction of external strain.  If a collagen gel is cut perpendicular
to the direction of tensile strain and if cells are present in
sufficient numbers, they round up and reorient parallel to the free
surface introduced \cite{c:taka96}.

The response of adherent animal cells to mechanical input has evolved
in the physiological context of a multicellular organism and plays a
crucial role in development, tissue maintenance, angiogenesis, wound
contraction, inflammation and metastasis. Recently, there has been a
large experimental effort to understand its molecular basis. A growing
body of evidence suggests that focal adhesions based on transmembrane
receptors from the integrin family act as mechanosensors which
directly feed into cellular regulation \cite{c:geig01a}. In
particular, the application of external force to focal adhesions leads
to their structural reinforcement and strong signalling activity
\cite{c:wang93,c:choq97,uss:rive01} and internally generated force
correlates with the state of aggregation of mature focal adhesions
\cite{uss:bala01,c:tan03}. The exact mechanism of the mechanosensor at focal
adhesions is still unknown, although structural reorganization of the
whole aggregate or conformational changes of specific molecules are
likely candidates. Although focal adhesions are characteristic for
cells cultured on flat and rigid substrates, cells in a soft
environment develop similar cell-matrix contacts which presumably have
the same mechanosensory function \cite{c:cuki02}.

As a result of active mechanosensing at cell-matrix contacts, cells
remodel their contacts and cytoskeleton. In particular, they might
change position and become oriented in a certain direction, depending
on the mechanical properties of their environment. Although cellular
behavior in principle results from very complex regulatory processes,
here we show that the typical cellular reaction to mechanical input
seems to be a simple preference for large effective stiffness:
starting from this principle, we are able to explain many experimental
findings which have been reported for the behavior of adherent cells
both on elastic substrates and in hydrogels. In order to make these
predictions, we have to calculate how stress and strain propagates in
the extracellular environment. For this purpose, we model it with
linear elasticity theory and solve the elastic equations for different
geometries and boundary conditions of interest. We then calculate
the position and orientation in which the cell senses maximal effective
stiffness in its local environment.  Predicting cell organization in a
soft medium not only contributes to a better understanding of many
physiological situations, but also is of large practical value for
application in tissue engineering, e.g.\ when culturing fibroblasts in
collagen gels. 

\section*{Theory}

\textbf{Optimization principle for single contact.} Motivated mainly
by recent experiments with elastic substrates
\cite{c:pelh97,c:lo00,c:wong03}, we suggest that an adherent cell
positions and orients itself in such a way that it senses maximal
effective stiffness in its environment.  In these experiments, the
most relevant input for cellular decision making is local elasticity
of the surrounding environment. Thus we first have to calculate how
stress and strain in the medium is propagated towards the cell. These
calculations are in general very complicated and will be presented
below for different situations of interest.  In order to keep our
calculations feasible, we assume that the extracellular environment is
described by isotropic linear elasticity theory. This holds true for
most synthetic elastic substrates and might be a reasonable assumption
for hydrogels.  Hence, there are two elastic moduli, the Young modulus
$E$ (which describes rigidity) and the Poisson ratio $\nu$ (which
describes the relative weight of compression and shear modes). In
practice, $E$ will be on the order of kPa, which is a typical
physiological value for tissue stiffness.  In most situations, $\nu$
is expected to be close to $1/2$ (the value for an incompressible
medium), but other values might be realized in future applications.

We then ask in which way the cell will organize itself if it probes
its local environment by actively pulling on it. In particular, we aim
to define a quantity which describes the kind of information which the
cell can extract from its soft environment with the help of its
contractile machinery. We first consider a single cell-matrix contact
and suggest that an appropriate choice is the work $W$ which the cell
has to invest into the surrounding elastic medium in order to build up
some force $\vec F$ at the contact position $\vec{r_c}$. As we will
show now, the quantity $W$ can be used to describe the effects of
increased rigidity $E$ and prestrain in the elastic environment on an
equal basis. Therefore $W$ is a measure for the effective stiffness of
the elastic environment as probed through a single contact.

In the absence of prestrain, the work $W$ invested into the environment is
\begin{equation}
W_0 = \frac{1}{2} \int d^3r\ C_{ijkl} u^c_{ij}(\vec r) u_{kl}^c(\vec
r) = \frac{1}{2} \vec F \cdot \vec u_c(\vec r_c),
\label{eq:Wc}
\end{equation}
where summation over repeated indices is implied. Here $\vec{u_c}$ is
the displacement caused by the cell, $u^c_{ij}(\vec r)$ the
corresponding strain tensor and $C_{ijkl}$ the elastic constant tensor
based on $E$ and $\nu$.  The volume integral runs over the whole space
filled with extracellular material and its conversion into a local
expression requires partial integration and use of the mechanical
equilibrium conditions (details of our calculations will be published
elsewhere). Formally, the self-energy of a given contact diverges for
a point force, but this divergence can easily be removed by assuming
distributed force. Since displacement decreases with increasing
rigidity ($u_{ij} \sim 1/E$), the cell has to invest less work $W_0$
in order to achieve a certain force $\vec F$ when rigidity $E$
increases. Hence, the cell senses maximal stiffness at the contact
when $W_0$ is minimal.

In a homogeneous medium, the elastic constants do not change and $W_0$
is a constant. However, the work $W$ needed to build up some force
$\vec F$ at the contact position $\vec{r_c}$ can vary due to the
presence of prestrain. The corresponding contribution to $W$ is
\begin{equation}
\Delta W = \int d^3r\ C_{ijkl} u_{ij}^c(\vec r) u_{kl}^e(\vec r) 
= \vec F \cdot \vec u_e(\vec r_c),
\label{eq:DeltaW}
\end{equation}
where $\vec{u_e}$ is the displacement caused by the external strain
and $u^e_{ij}(\vec r)$ the corresponding strain tensor. Since a
negative $\Delta W$ reduces the cellular work $W = W_0 + \Delta W$, as
does a larger rigidity $E$, it represents an effective stiffening of
the environment (\textit{strain-stiffening}). Correspondingly, a
positive $\Delta W$ represents an effective softening with respect to
the unstrained medium. Therefore the quantity $W$ allows to
characterize the local elastic input available to an actively
mechanosensing cell within the unifying concept of effective
stiffness, independent of its physical origin, which might be rigidity
or prestrain. In the following, we will identify optimal cell position
and orientation with the specific force pattern which minimizes the
quantity $W$. In the sense described here, this corresponds to a
cellular preference for maximal effective stiffness in its local
elastic environment.

It is important to note that conceptually the principle suggested here
does not imply that the cell actually minimizes the work $W$ invested
into its soft environment. Instead we suggest here that calculating
the quantity $W$ for different situations of interest is an
appropriate measure for the kind of information a cell can extract
from its elastic environment through active mechanosensing. The real
justification of our model will be its success in explaining a large
body of experimental data (see results section). Nevertheless, below
we will also present some potential mechanism for the cellular
preference for effective stiffness, which in fact uses the quantity
$W$ not as a characterization of the external environment, but as a
relevant quantity for some internal mechanism.

\textbf{Optimization principle for cellular force pattern.} Different 
contacts are coupled through the actin cytoskeleton in such a way that
overall force balance is ensured.  We account for this constraint by
considering only pairs of opposing forces.  In elasticity theory, such
a pinching force pattern is known as an \textit{anisotropic force
contraction dipole}, that is the tensor $P_{ij} = P n_i n_j$, where
$P$ is the dipole strength, the product of force magnitude and force
separation, and $\vec n$ its orientation
\cite{e:siem68,e:wagn74,uss:schw02a}. Typical cellular dipoles
have been measured to be of the order of $P \approx - 10^{-11} J$
(this corresponds to two forces of 200 nN each, separated by a
distance of 60 $\mu$m) \cite{uss:schw02b}. The effect of external
strain on the work required to build up the force dipole $P_{ij}$ at
the cell position $\vec r_c$ can be written as
\begin{equation}
\Delta W = P_{ij} u^e_{ij}({\vec r_c}).
\label{eq:DeltaW2}
\end{equation}
Like in the case of a single contact, optimal cell organization can be
identified with the specific force pattern which minimizes $\Delta
W$. It follows directly from \eq{eq:DeltaW2} that due to the
contractile activity of the cell ($P < 0$), tensile strain ($u^e_{ij}
> 0$) will always be favorable (negative $\Delta W$).  In contrast to
compressive strain ($u^e_{ij} < 0$), it corresponds to an effective
increase in stiffness. Note that in contrast to position, orientation
and sign, the magnitude $|P|$ of the cellular force pattern does not
matter in the model presented here. This reflects the fact that here
we aim to characterize the mechanical properties of the extracellular
environment sensed by the cell, rather than the process of active
mechanosensing itself.

\begin{figure}
\begin{center}
\includegraphics[width=\columnwidth]{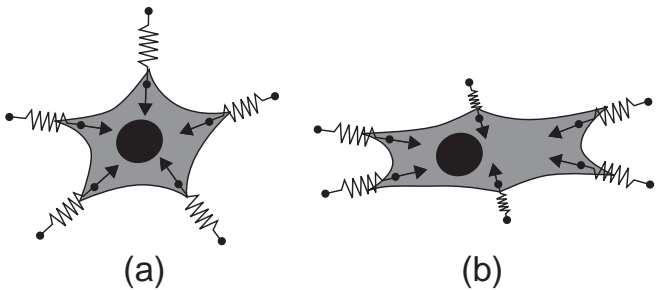}
\caption{\label{FigFocalContacts}An adherent cell actively pulls on its soft 
environment through cell-matrix contacts. Experimentally, one finds
that cells orient themselves in the direction of maximal stiffness of
the environment. With this cartoon, we present one possible mechanism
by which active mechanosensing in an elastically anisotropic medium
might lead to cell orientation. The local elastic environment is
represented by linear springs with different spring constants $K$, as
indicated by differently sized springs. For upregulation of a contact,
the cell has to invest the work $F^2/2K$.  Therefore, upregulation is
more efficient for larger $K$. (a) In an isotropic environment, all
spring constants are the same, growth at different contacts is similar
and the cell does not orient.  (b) If spring constants are largest
in one specific direction, the corresponding contacts outgrow the
others and the cell orients in the direction of maximal stiffness of
the environment. In this paper, we use the cellular preference for
large effective stiffness and modelling of the extracellular
environment by linear elasticity theory to predict cell positioning
and orientation in soft media.}
\end{center}
\end{figure}

\textbf{Possible origin of optimization principle.}  Our modelling
starts from the phenomenological observation that cells seem to prefer
maximal effective stiffness in their environment.  Although it can be
justified by its large success in explaining experimental observations
(see results section), we also want to suggest a possible mechanism
for our main assumption.  For this purpose, we use a simple
one-dimensional analogue.  Consider the extracellular environment to
act like a linear spring with spring constant $K$, on which the cell
is pulling through a single cell-matrix contact. Recent experiments on
focal adhesions \cite{uss:rive01} suggest that upregulation of contact
growth is related to reaching a certain threshold in force $F$,
although the details of how force affects regulation are just
beginning to emerge \cite{c:geig01a}. In order to build up sufficiently
large force $F$, the cell has to invest energy $W=F^2/2K$ into the
spring. Thus, the stiffer the spring (the larger $K$), the less work
is needed and the more efficient the build-up of force will be. An
equivalent viewpoint is to assume that the cell invests the power $L$
into stretching the spring. Then it takes the time $t=F^2/2KL$ to
reach the force $F$. Therefore, a specific contact will grow faster
than the other contacts if it encounters a larger stiffness $K$.  In
principle, the cellular program could also be geared towards achieving
a certain displacement of the surrounding material, which would result
in a preference for effective softness of the environment. However,
this scenario would imply the existence of some additional mechanism
for outside-in signaling.  It is more realistic to consider that force
activates the cellular response through a certain displacement of
elastic components located \textit {inside} the cell. Since internal
displacement and force are expected to be linearly related through
another (internal) spring constant, one arrives at the same result.

Adhering cells probe the mechanical properties of their environment by
pulling at many cell-matrix contacts simultaneously
(\fig{FigFocalContacts}).  At each newly formed contact the cell is
expected to pull with a similar investment of resources (e.g.\ actin,
myosin or ATP).  However, each contact encounters a different elastic
environment, each of which can be represented by a different spring
constant. In an isotropic situation (\fig{FigFocalContacts}a), all
spring constants are equal, the contacts have similar growth behavior
and there is no reason for a cell to orient. Experimentally, the
cell adopts a round or stellate morphology, depending on the number of
cell-matrix contacts. In an anisotropic situation
(\fig{FigFocalContacts}b), build-up of force is more efficient in one
specific direction and contacts in this direction will eventually
outgrow the other ones.  Here, the anisotropic elastic properties of
the medium provide an orientational clue for the cell, which orients
along the direction of maximal effective stiffness. Depending on e.g.\
the presence of motility factors, this orientation response might be
followed by cell locomotion.

\section*{Results}

\textbf{Homogeneous external strain}. We first consider a cell interacting
with homogeneous external strain, either on the top surface of a
rectangular slab of elastic material (elastic substrate) or inside an
infinite elastic material (hydrogel). In both cases, the equations
of three-dimensional isotropic elasticity give
\begin{equation}
\Delta W = - \frac{P p}{E} \left[ (1+\nu) \cos^2{\theta} - \nu \right]
\label{eq:strain}
\end{equation}
where $\theta$ is the orientation angle relative to the direction of
the externally applied tensile stress $p < 0$. Optimal cell
orientation corresponds to minimal $\Delta W$, which is achieved for
$\theta = 0$, irrespective of the Poisson ratio $\nu$.  Thus the cell
orients preferentially with the direction of stretch. This behavior is
indeed observed experimentally, both for fibroblast on elastic
substrates \cite{c:hast83} and in collagen gels
\cite{c:bell79,c:east98}. Since $\Delta W$ decreases with increasing
rigidity $E$, the elastic effects discussed here will be observed only
in a soft environment, namely with rigidity $E$ around kPa, which is a
typical physiological value for tissue stiffness. For stiffer
substrates the variations in $\Delta W$ for different contact
positions might become too small to induce an orientation response.

\begin{figure}
\begin{center}
\includegraphics[width=\columnwidth]{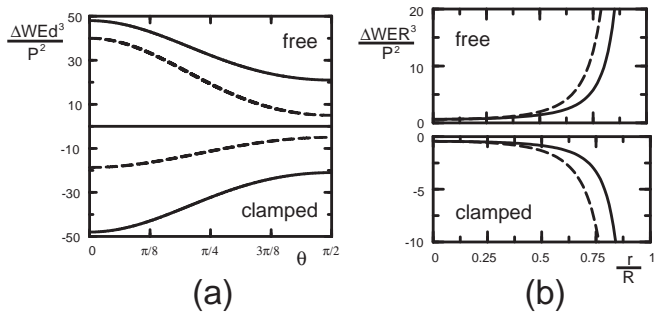}
\caption{\label{Fig3}Adjusting cell position and orientation in such
a way that the cell senses maximal effective stiffness in its
environment is equivalent to minimizing the quantity $W$, the
amount of work the cell invests into the elastic surroundings in the
presence of external strain. In the presence of mechanical activity,
sample boundaries induce external strain which can result in different
cell organization. (a) $\Delta W$ for a cell with dipole strength $P$
which is a distance $d$ away from the surface of an elastic halfspace
with rigidity $E$, plotted in units of $P^2/E d^3$ as a function of
angle $\theta$ between cell orientation and surface normal (rescaled
by $256 \pi$). Solid and dashed lines correspond to Poisson ratios
$\nu = 1/2$ and $\nu = 0$, respectively. Irrespective of $\nu$, the
optimal orientations (minimal $\Delta W$) are perpendicular ($\theta =
0$) and parallel ($\theta = \pi/2$) to the surface for clamped and
free boundaries, respectively.  Since $|\Delta W|$ increases if $d$
decreases, the overall mechanical activity of a cell increases towards
a clamped surface ($\Delta W < 0$), but decreases towards a free
surface ($\Delta W > 0$). (b) $\Delta W$ for a cell in an elastic
sphere of radius $R$, plotted in units of $P/E R^3$ as a function of
distance $r$ to the sphere center in units of $R$ for $\nu = 1/3$
(rescaled by $15/8$). Solid and dashed lines are parallel ($\theta =
\pi/2$) and perpendicular ($\theta = 0$) orientations, respectively
(all other orientations yields curves which lie inbetween the ones
shown). Like in an elastic halfspace, parallel and perpendicular
orientations are favored (minimal $\Delta W$) for free and clamped
boundaries, respectively. For clamped boundaries, mechanical activity
is favored (smaller $\Delta W$) towards the surface. For free
boundaries, mechanical activity is disfavored (larger $\Delta W$)
towards the surface.}
\end{center}
\end{figure}

\begin{figure}
\begin{center}
\includegraphics[width=\columnwidth]{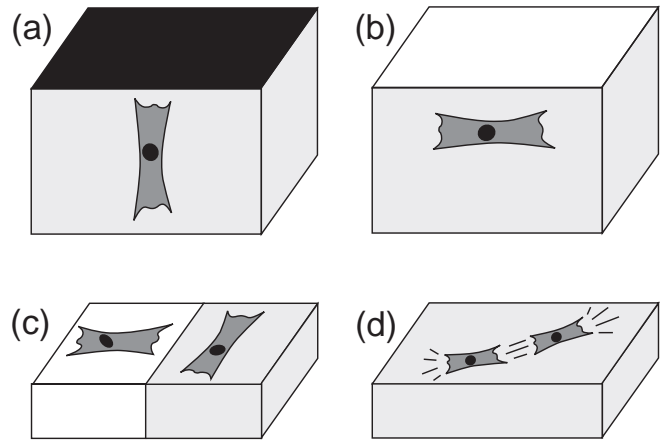}
\caption{\label{Fig2}Predicted cell orientation in a hydrogel close
to a surface (a,b) and on elastic substrates (c,d). (a) Cells prefer
the direction of maximal effective stiffness. Thus, they orient
perpendicular to a clamped surface. (b) For a free surface, this
direction is parallel to the surface.  (c) Cells close to a boundary
between soft (left) and rigid (right) regions prefer analogous
orientations as cells close to clamped and free surfaces in a hydrogel,
respectively.  (d) Cells interact elastically to form strings, because
in nose-to-tail alignment, the mechanical activity of one cell
triggers the one of the other cell, thereby forming a positive
feedback loop.}
\end{center}
\end{figure}

\textbf{Boundaries}. In a physiological context, cells are often close 
to boundaries, like the surface of a tissue or organ.  In the presence
of cell traction, boundaries alter the strain with respect to a
homogeneous infinite medium by a boundary-induced strain
(\textit{image strain}), which has a similar effect like external
strain. In this way, cells can actively sense not only the presence of a
close-by surface, but also its shape and boundary conditions. In order
to predict the effect of boundaries on cell organization, we study a
semi-infinite space with a planar surface, for which the elastic
equations can be solved exactly
\cite{e:walp96}. The details of the boundary conditions in a
physiological context can be very complicated.  Here we address 
two fundamental reference cases, namely free and clamped boundary
conditions, for which normal stress and displacement, respectively,
vanish at the surface. Consider a force dipole which is a distance $d$
away from the planar surface and has an angle of orientation $\theta$
relative to the surface normal.  We find
\begin{equation}
\label{eq:calchalfspace}
\Delta W=\frac{P^2}{Ed^3}(a_{\nu}+b_{\nu} \cos^2 \theta+c_{\nu} \cos^4\theta),
\end{equation}
where the coefficients $a_{\nu}$, $b_{\nu}$ and $c_{\nu}$ are
complicated functions of $\nu$. $\Delta W$ scales quadratically with $P$
because the image strain scales linearly with $P$ (in other words, the
force dipole interacts with its images). For free and clamped
surfaces, the coefficients are positive and negative, respectively,
irrespective of $\nu$.  Therefore, the optimal configurations (minimal
$\Delta W$) are parallel ($\theta = \pi/2$) and perpendicular ($\theta
= 0$) for free and clamped boundaries, respectively, as plotted in
\fig{Fig3}a.  A schematic representation (\fig{Fig2}a and b) provides
a simple interpretation: for clamped (free) boundary conditions, the
cell senses maximal stiffness perpendicular (parallel) to the
surface. One may think of a clamped (free) surface as the interface
between the medium and a imaginary medium of infinite (vanishing)
rigidity, which effectively rigidifies (softens) the medium towards
the boundary.  In general, we find that free and clamped boundary
conditions always have opposite effects, albeit with one essential
difference: for clamped boundaries, mechanical activity of cells is
favored and cells can amplify this effect by adjusting
orientation. For free boundaries, mechanical activity of cells is
disfavored and the orientation response is an aversion response.

Experimentally, it is well known that mechanical activity of cells
increases for clamped boundary conditions \cite{c:grin00}.  The
predicted orientation effects close to boundaries have been observed
numerous times, e.g.\ the parallel orientation of cells close to free
surfaces \cite{c:bell79}. Our model predicts the same orientation
effects for an elastic substrate with two regions of different
rigidities (\fig{Fig2}c): cells on the soft and stiff sides of the
boundary orient perpendicular and parallel to it, respectively.
Indeed fibroblasts migrating from a soft to a stiff region keep their
perpendicular orientation and cross over to the stiff side, while
fibroblasts migrating from a stiff to a soft region do not cross the
boundary, but turn by 90 degrees and move parallel to the boundary
\cite{c:lo00}.

\textbf{Finite sized sample.} As an example for a finite sized sample, 
we consider the elastic sphere with radius $R$. The elastic equations
can be solved exactly by using an expansion in terms of vector spherical
harmonics \cite{e:hirs81}. We find
\begin{equation} \label{eq:sphere}
\Delta W=\frac{P^2}{E R^3} f_{\nu}(\frac{r}{R},\theta),
\end{equation}
where $r$ denotes distance to the sphere center, $\theta$ the
orientation in respect to the radial direction and $f_{\nu}$ is an
infinite sum over all angular momenta, which does not change
qualitatively as $\nu$ is varied. In regard to orientation, we find
the same results as for the elastic halfspace (compare \fig{Fig3}b):
cells will orient parallel (perpendicular) to free (clamped) surfaces,
respectively.  We also find a similar result for the effect of
distance to the surface: for free (clamped) boundary conditions, a
small (large) distance to the sphere center is more favorable, since
the surface favors (disfavors) mechanical activity. The new aspect
here is the role of sphere radius $R$: since $|\Delta W|$ increases
when sphere radius $R$ decreases, one can effectively rigidify
(soften) a material with clamped (free) boundaries by reducing system
size. Our predictions could be tested using e.g.\ fibroblast-populated
collagen microspheres, an assay which has been introduced to study
compaction of tissue equivalents at high cell density
\cite{c:moon93}. Since here we are mainly concerned with single cell
effects, we suggest to modify this assay in such a way as to monitor
the organization of isolated cells close to the sphere surface at low
cell density and as a function of varying sphere radius.

\begin{figure}
\begin{center}
\includegraphics[width=0.5\columnwidth]{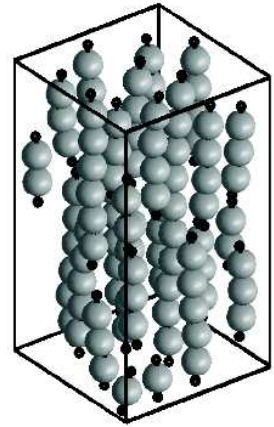}
\caption{\label{FigMonteCarlo}Monte Carlo simulations of elastically
interacting cells in an external strain field. The temperature used in
the simulation represents the stochastic element of the process of
cell organization. Without external strain, cells form strings. In its
presence, strings align in parallel.}
\end{center}
\end{figure}

\textbf{Cooperative effects.} Up to now we have been discussing
single cell effects and we now turn to cooperative effects.  In
particular, we now consider the case that external strain is caused by
the traction of other cells, which amounts to an elastic interaction
of cells. Even if all cells initially have isotropic force patterns,
they will sense anisotropic strain fields and start to orient
themselves.  For the simplest case of two cells, we find
\begin{equation}
\Delta W=\frac{P^2}{E r^3} g_{\nu}(\theta_{1},\theta_{2},\theta),
\end{equation}
where $r$ is the distance between the force dipoles and $g_{\nu}$ is a
complicated function of $\nu$ and the three orientational degrees of
freedom. Our calculation shows that $\Delta W$ has a pronounced
minimum for completely aligned dipoles, independent of $\nu$
(\fig{Fig2}d). In this configuration, both cells sense maximal
effective stiffness, because maximal strain stiffening occurs along
the axis of contraction.  This finding suggests that a common pattern
for the organization of elastically interacting cells will be the
formation of strings of cells. They might close into rings, so that
each cell can be fully activated by its two neighbors.  We used Monte
Carlo simulations to obtain a typical configuration of elastically
interacting cells in an external strain field (\fig{FigMonteCarlo}).
The temperature of the Monte Carlo simulation represents the
stochastic nature of the orientation response.  We find strings of
cells aligned in parallel with the external strain, exactly as
observed experimentally \cite{c:east98}.

It is important to note that there is a positive feedback for cell
alignment: the more cells orient in one direction, the stronger
becomes the input for other cells to adopt the same orientation.  For
example, it has been reported that when boundary condition are changed
from clamped to free by cutting the collagen gel, fibroblasts show the
predicted reorientation by 90 degrees only when sufficiently many
cells are present \cite{c:taka96}. In fact our calculations and
simulations show that cells can orient in parallel even with respect
to clamped boundaries, if there are sufficiently many cells such that
the direct elastic interaction between cells dominates the single cell
response of perpendicular orientation.  In practise, the single cell
response might also be disturbed because elastic signals could be
screened by traction of randomly oriented cells.  Indeed such an
effect has been reported for experiments with elastic substrates
\cite{c:lo00}.

\section*{Discussion}

It has long been known, especially in the medical and bioengineering
communities, that cell organization in soft media is strongly
influenced by the mechanical properties of the environment.  Here we
presented a model which is able to explain numerous experimental
observations that have been reported for organization of cells
(especially fibroblasts) both on elastic substrates and in hydrogels.
The excellent agreement of our results with experiments suggests that
cell organization can be predicted from local mechanical properties
which the cell actively senses in its environment. In fact the only
property of cellular regulation which enters our model is the
assumption that cells locally prefer large effective stiffness.
Otherwise our modelling focuses on the elastic properties of the
extracellular environment. 

Modelling the soft environment of cells as an isotropic elastic medium
is certainly a good assumption for elastic substrates. The situation
is more complicated for hydrogels, in particular because they might
not behave elastically and because they feature fiber degrees of
freedom. Cell organization in gels is often explained
by contact guidance, the alignment of cells along topographic features
like collagen fibers. Since fibers can become aligned due to cell
traction, contact guidance provides a long-ranged and persistent
mechanism for cellular self-organization in tissue equivalents
\cite{c:oste83}. This process has been modeled before. 
In the theory of Ref.~\cite{c:oste83}, flux equations for
cellular and matrix densities are combined with mechanical equations
which include cells as centers of isotropic contraction. This might be
a good model for chondrocytes, which tend to show a spherical
morphology. The anisotropic biphasic theory (ABT) from
Ref.~\cite{c:baro97} aims at cells like fibroblasts and
smooth muscle cells, whose typical morphology in tissue equivalents is
bipolar. ABT introduces a cell orientation tensor, which is coupled to
a fiber orientation tensor, since cells are assumed to react
foremost to fiber degrees of freedom. In our model, the force dipole
tensor represents cell orientation as does the cell orientation tensor
in ABT, but it is coupled to elastic degrees of freedom, since cells
are assumed to react foremost to large effective stiffness.

Because models for contact guidance in tissue equivalents focus on fiber
degrees of freedom and high cell densities, they do not explain the
single cell responses observed on elastic substrates, where contact
guidance usually is ruled out \cite{c:hast83,c:lo00}.  The large
predictive power of our model for elastic substrate experiments
suggests that active mechanosensing by single cells might also be
involved with cell organization in hydrogels. However, for the
collagen assay from Ref.~\cite{c:girt02} it has been shown that as a
result of external strain, fibers become rearranged and stress
relaxes towards zero.  In a matrix which cannot support any stress,
our elastic considerations do not apply and contact guidance through
formerly aligned fibers might be the only relevant clue for cell
organization \cite{c:girt02}. However, it is important to note that in
our model, stress is actively generated by cells and thus needs to be
supported only over time scales in which the cell actively senses the
mechanical properties of its environment. In particular, if fiber
alignment has resulted in some anisotropic mechanical environment, the
cell might sense the anisotropic mechanical properties of the matrix and
orient itself correspondingly. This might explain why cells have been
found to align to a greater extent with respect to external strain
than the surrounding collagen fibrils \cite{c:girt02} and why our
modelling is also successful for hydrogels.  In general, future
experiments are needed to clarify the relative importance of
topographic versus mechanical clues for cell organization in
hydrogels, while future modelling is needed to account for the
mechanical (in particular, viscoelastic) properties of hydrogels.

We also want to point out that contact guidance is a bidirectional
clue and provides only guidance, in contrast to external elastic
strain, which provides taxis. In our model, taxis is reflected by the
position dependence of $\Delta W$. For example, our theory not only
predicts that cells prefer to orient parallel to free boundaries, but
also that cells prefer to move away from them.  Moreover a simple
preference for cell alignment along fibers does not predict what cells
do if they encounter a fiber junction in the gel.  Although we are not
concerned with cell locomotion here, our modelling would suggest that
cells prefer the fiber under largest tension, exactly as has been
observed experimentally for neutrophils migrating in human amnion
\cite{c:mand97}. 

During recent years, the regulated response to mechanical input by
single cells has been studied experimentally in larger detail.  There
is a growing body of evidence now that integrin-based cell-matrix
contacts act as local mechanosensors which channel mechanical
information about the environment directly into cellular decision
making. Although this does not concern our modelling directly, here we
suggested that the upregulation of growth of cell-matrix contacts in a
stiff environment might result from the fact that it is triggered by a
threshold in force, whose build-up is more efficient for larger
stiffness. An equivalent viewpoint is that growth of cell-matrix
contacts is faster on stiffer substrates. As experimental test for this
hypothesis, we suggest correlation studies for growth of cell-matrix
contacts and cellular organization, especially close to sample
boundaries, where cells can amplify the mechanical input provided by
boundary induced strain through active mechanosensing. Quantitative
data about growth behavior of cell matrix contact will allow us to
further refine our model in a more quantitative way, possibly also
including modelling of cellular features like morphology and force
pattern, which are not the focus of this work.
 
It is important to note that our model suggests completely different
behavior for cells then one would expect for physically inert
particles interacting with a soft matrix (like mobile inclusions in a
metal). Although the expression in \eq{eq:DeltaW2} is similar to the
interaction potential for the physical system of force dipoles in an
external strain field, it has the opposite sign, because in the
physical system, one has to minimize the composite energy of defect
and medium \cite {e:siem68,e:wagn74}. The physical potential has been
used before to model elastic interactions of cells without any
regulatory response and has been shown to lead to aggregation behavior
similar to the one of electric quadrupoles \cite {uss:schw02a}, while
the model introduced here leads to aggregation behavior similar to the
one of electric dipoles \cite{e:tlus00}.  

Since elastic effects are long-ranged and propagate quickly, they
provide an appealing mechanism for signal transduction for
mechanically active cells in soft media.  However, they are also
unspecific and cells might not be able to distinguish between
different sources. On the other hand, additional information channels,
like soluble ligands, will certainly supplement elastic signals.
Moreover, cells in highly differentiated organisms are likely to
interprete mechanical signals only in their own physiological context,
which is more restricted than for cells in an arbitrary environment.

In summary, we have presented an optimization principle in linear
elasticity theory that allows to predict cell organization in soft
media in excellent agreement with a large body of experiments.
Moreover we have suggested a mechanism which links the cellular
preference for large effective stiffness to growth of cell-matrix
contacts. Our modelling results in many interesting predictions which
now can be checked experimentally. In the future, our model might be
extended to high cell densities and strong cooperative effects, which
are characteristic for tissue equivalents. We expect that then it can
be used for rational design in tissue engineering. For example, using
numerical (finite element) methods, one can use it to optimize
protocols for the design of tissue equivalents for implants in regard
to geometry and boundary conditions.

We thank Martin Bastmeyer, Benjamin Geiger, Dirk Lehnert, Rudolf
Merkel and Samuel Safran for critical reading of an earlier version of
the manuscript. We also thank the anonymous referees for helpful
suggestions.  This work was supported by the Emmy Noether Program of
the German Science Foundation.


\begin{thebibliography}{10}

\bibitem{c:galb98}
Galbraith, C.~G. \& Sheetz, M. (1998) {\em Curr. Opin. Cell Biol.} {\bf 10},
  566--571.

\bibitem{c:huan99}
Huang, S. \& Ingber, D.~E. (1999) {\em Nat. Cell Biol.} {\bf 1}, E131--E138.

\bibitem{c:geig02}
Geiger, B. \& Bershadsky, A. (2002) {\em Cell} {\bf 110}, 139--142.

\bibitem{c:harr80}
Harris, A.~K., Wild, P., \& Stopak, D. (1980) {\em Science} {\bf 208},
  177--179.

\bibitem{c:pelh97}
Pelham, R.~J. \& Wang, Y.-L. (1997) {\em Proc. Natl. Acad. Sci. USA} {\bf 94},
  13661--13665.

\bibitem{c:hast83}
Haston, W.~S., Shields, J.~M., \& Wilkinson, P.~C. (1983) {\em Exp. Cell Res.}
  {\bf 146}, 117--126.

\bibitem{c:lo00}
Lo, C.-M., Wang, H.-B., Dembo, M., \& Wang, Y.-L. (2000) {\em Biophys. J.} {\bf
  79}, 144--152.

\bibitem{c:wong03}
Wong, J. Y., Velasco, A., Rajagopalan, P., \& Pham, Q. (2003) {\em Langmuir} {\bf
  19}, 1908--1913.

\bibitem{c:bell79}
Bell, E., Ivarsson, B., \& Merrill, C. (1979) {\em Proc. Natl. Acad. Sci. USA} {\bf 76}, 1274--1278.

\bibitem{c:east98}
Eastwood, M., Mudera, V.~C., Mc{G}routher, D.~A., \& Brown, R.~A. (1998) {\em
  Cell Motil. Cytoskel.} {\bf 40}, 13--21.

\bibitem{c:taka96}
Takakuda, K. \& Miyairi, H. (1996) {\em Biomaterials} {\bf 17}, 1393--1397.

\bibitem{c:geig01a}
Geiger, B., Bershadsky, A., Pankov, R., \& Yamada, K.M. (2001) {\em Nat. Rev.
  Mol. Cell Biol.} {\bf 2}, 793--805.

\bibitem{c:wang93}
Wang, N., Butler, J.~P., \& Ingber, D.~E. (1993) {\em Science} {\bf 260},
  1124--1127.

\bibitem{c:choq97}
Choquet, D., Felsenfeld, D.~F., \& Sheetz, M.~P. (1997) {\em Cell} {\bf 88},
  39--48.

\bibitem{uss:rive01}
Riveline, D., Zamir, E., Balaban, N.~Q., Schwarz, U.~S., Geiger, B., Kam, Z.,
  \& Bershadsky, A.~D. (2001) {\em J. Cell Biol.} {\bf 153}, 1175--1185.

\bibitem{uss:bala01}
Balaban, N.~Q., Schwarz, U.~S., Riveline, D., Goichberg, P., Tzur, G., Sabanay,
  I., Mahalu, D., Safran, S., Bershadsky, A., Addadi, L., \& Geiger, B. (2001)
  {\em Nat. Cell Biol.} {\bf 3}, 466--472.

\bibitem{c:tan03}
Tan, J.~L., Tien, J., Pirone, D.~M., Gray, D.~S., Bhadriraju, K., \& Chen, C. S. (2003)
{\em Proc. Natl. Acad. Sci. USA} {\bf 100}, 1484--1489.

\bibitem{c:cuki02}
Cukierman, E., Pankov, R., Stevens, D.R., \& Yamada, K.M. (2002) {\em Science}
  {\bf 294}, 1708--1712.

\bibitem{e:siem68}
Siems, R. (1968) {\em Phys. Stat. Sol.} {\bf 30}, 645--658.

\bibitem{e:wagn74}
Wagner, H. \& Horner, H. (1974) {\em Adv. Phys.} {\bf 23}, 587.

\bibitem{uss:schw02a}
Schwarz, U.~S. \& Safran, S.~A. (2002) {\em Phys. Rev. Lett.} {\bf 88}, 048102.

\bibitem{uss:schw02b}
Schwarz, U.~S., Balaban, N.~Q., Riveline, D., Bershadsky, A., Geiger, B., \&
  Safran, S.~A. (2002) {\em Biophys. J.} {\bf 83}, 1380--1394.

\bibitem{e:walp96}
Walpole, L.~J. (1996) {\em Int. J. Engng. Sci.} {\bf 34}, 629--638.

\bibitem{c:grin00}
Grinnell, F. (2000) {\em Trends in Cell Biol.} {\bf 10}, 362--365.

\bibitem{e:hirs81}
Hirsekorn, R.-P. \& Siems, R. (1981) {\em Z. Phys. B - Cond. Mat.} {\bf 40},
  311--319.

\bibitem{c:moon93}
Moon, A.~G., and Tranquillo, R.~T. (1993)
{\em AIChE J.} {\bf 39}, 163--177.

\bibitem{c:oste83}
Oster, G.~F., Murray, J.~D., \& Harris, A.~K. (1983) {\em J. Embryol. Exp.
  Morph.} {\bf 78}, 83--125.

\bibitem{c:baro97}
Barocas, V.~H., and Tranquillo, R.~T. (1997)
{\em J. Biomech. Eng.} {\bf 119}, 137--145.

\bibitem{c:girt02}
Girton, T.~S., Barocas, V.~H., and Tranquillo, R.~T. (2002)
{\em J. Biomech. Eng.} {\bf 124}, 568--575.

\bibitem{c:mand97}
Mandeville, J.~T.~H., Lawson, M.~A., \& Maxfield, F.~R. (1997) {\em J. Leukoc.
  Biol.} {\bf 61}, 188--200.

\bibitem{e:tlus00}
Tlusty, T. \& Safran, S.~A. (2000) {\em Science} {\bf 290}, 1328--1331.

\end{thebibliography}

\end{document}